\documentstyle[epsfig,12pt]{article}

\begin{document}

%\draft   
%\large\normalsize
\begin{center}
{\Large \bf Procedure for direct measurement of the 
\bigskip

 Cabibbo-Kobayashi-Maskawa angle $\gamma$}

\vskip 1.5 cm
{\large I. Bediaga$^{a}$, R.E. Blanco$^{b}$, C. G\"obel$^{a,b}$, 
and R. M\'endez--Galain$^{b}$ }

\vskip 1 cm
{\it \small a) Centro Brasileiro de Pesquisas F\'\i sicas, 
R. Dr. Xavier Sigaud 150, 

 22290 -- 180 --  Rio de Janeiro, RJ, Brazil

 b) Instituto de F\'{\i}sica, Facultad de Ingenier\'{\i}a,
CC 30, CP 11000 Montevideo, Uruguay}
\end{center}

\vskip 1.5 cm
\begin{abstract}

A natural procedure is presented to measure the angle $\gamma$ from the 
decay $B^{\pm} \to \pi^{\pm}\pi^+\pi^-$. It is based in the Dalitz plot 
fitting analysis. Neither amplitudes nor strong phases have to be known 
a priori.  We present simulations of this decay computing both 
statistical and theoretical uncertainties and analyze the experimental 
feasibility. We found that $\gamma$ could be measured with a combined 
error of the order of 20$^o$ 
with 90\% of CL after about a couple of years
of running of the first generation of B factories.

\end{abstract}

%\twocolumn

\vfill $^*$ To appear in Phys. Rev. Lett. {\bf 81} 4067 (1998).

\newpage

The study of hadronic decays in the B system seems to be a powerful 
tool for the understanding of CP violation. To check Standard Model 
(SM) predictions it is particularly important to  measure the three
Cabibbo-Kobayashi-Maskawa  (CKM) angles $\alpha\equiv 
arg({-V_{td}V^*_{tb}/V_{ud}V^*_{ub}}), 
\beta\equiv arg({-V_{cd}V^*_{cb}/V_{td}V^*_{tb}})$ and $\gamma\equiv 
arg({-V_{ud}V^*_{ub}/V_{cd}V^*_{cb}})$.
Only $\beta$ is expected to be clearly measured from the gold plated  
$B_0 \to J/\psi K_s$ decay which is almost free from theoretical 
uncertainties\cite{review} and benefits from the large ($10^{-3}$) 
branching ratio. The extraction of the two other angles requires the 
measurement of decays with branching ratios of the order of $10^{-5}$ 
or less. Many interesting methods using various decays have been 
proposed so far in the literature\cite{pap59,pap8,pap1} 
but the matter is still open. 

More precisely, the angle $\gamma$ seems to be hard to measure. The 
method presented in Ref. \cite{pap59} provides a theoretically clean 
procedure to extract 
$\gamma$ combining decays with $D^0$ in the final state. Unfortunately, 
both the original method and clever extensions\cite{pap8} of it demand a 
large statistics. As a result, one expects to need about ten 
years\cite{BaBar} 
of data taking in the first generation of B factories to attain a reasonable 
error --- at least 15$^o$ --- in the measurement of $\gamma$. It is then 
interesting to look for other methods that could provide a 
constraint for the value of $\gamma$ in much less time --- their eventual 
theoretical errors should be as well estimated as possible\cite{fleisher}.

In this letter we present a direct and simple method that could provide a 
nice first measurement of the angle $\gamma$ after about a couple of years
of data 
taking in the first generation of B factories. We use the decay 
$B^{\pm} \to \pi^{\pm}\pi^+\pi^-$ where the necessary interference is 
given by the intermediate resonant channel ${\chi_c}_0\pi^\pm$. This channel 
has been first pointed out in Ref. \cite{pap3}; nevertheless, in that 
reference 
the method used to extract the angle was very model dependent and demanded 
large statistics. Here, we present a totally different approach: we show the 
viability of performing a full Dalitz plot analysis of this decay. It can 
provide a {\it direct} measurement of the angle $\gamma$ free from model 
dependencies.

Other methods existing in the literature to measure 
$\gamma$, independently of the considered channel, are 
based in the measurement of {\it branching ratios} and 
asymmetries\cite{review}.  
The relationship between these measured {\it numbers} and the angle 
$\gamma$ 
is not direct. Moreover, these 
methods
 generally present discrete ambiguities.

The main feature of our method  
is that it exploits the fact that in three body decays 
one can have a direct measurement of the {\it amplitude} of a 
decay --- instead of branching ratios, that is amplitude squared. This means 
that one can have a {\it direct experimental access to the phase} of a given 
decay. This fact has already been used\cite{letter-quinn} in connection 
with CP violation, in a quite different context. 
The method presented in this letter can eventually be also used to extract 
CP violating angles from  other three body decays of charged B's.

Let us present our ideas using the channel $B^{+} \to \pi^{+}\pi^+\pi^-$.
Many intermediate channels contribute. Indeed, resonant channels --- 
$\rho^0\pi^+$, $f_0\pi^+$, ${\chi_c}_0\pi^+$, etc --- together with the direct 
non resonant decay produce the same experimentally detected final state. This 
final state is thus the product of the {\it interference} of all these 
intermediate states.

The fact that in three body decays one can measure differential widths --- 
usually displayed in a Dalitz plot (DP) --- allows a clean separation of  
these partial channels. The distribution of measured events in the plot can 
be fitted using appropriate fitting functions.

The fitting technique has proven to be very 
successful in describing, for example, three body decays of D 
mesons\cite{D-decays},  {\it even with only about one hundred of 
reconstructed events} \cite{Ds3pi}. 

In order to do so, one considers a fitting function including one term 
for each intermediate channel contributing to the final state. For 
example, for the decay 
$B^{+} \to \pi^{+}\pi^+\pi^-$ the fitting function should be
\begin{equation}
{{\cal F}_{B^{+} \to \pi^{+}\pi^+\pi^-}}(m^2_{1},m^2_{2}) = | \Sigma_i a_i 
e^{i\theta_i} F_i(m^2_{1},m^2_{2}) |^2 \hskip 0.1 cm ,
\label{fit}
\end{equation}
where $m^2_{1}=(p_{\pi^+_1}+p_{\pi^-})^2$ and 
$m^2_{2}=(p_{\pi^+_2}+p_{\pi^-})^2$ 
are the usual Dalitz plot invariant variables, $F_i$ are the amplitudes 
corresponding to each partial channel and $a_i$ and $\theta_i$ are unknown real 
parameters that will emerge from the fit. The sum is performed over all the 
intermediate resonances as well as the non-resonant decay. For the resonant 
channels, the function $F_i$ is very well known: it is simply the usual 
Breit-Wigner \cite{jackson} times an angular function according to the spin of 
the resonance. The non-resonant decay amplitude is discussed in 
\cite{prl-dalitz}; we will be back to it later.

The Dalitz plot maximum-likelihood technique uses the function of Eq. 
(\ref{fit}) to fit the measured differential width distribution 
${d\Gamma}/{dm^2_{1}dm^2_{2}}$ of the total decay. The output is the 
amplitude fractions $a_i$ {\it and phases $\theta_i$} of each partial decay. 
In other words, it brings a {\it direct measurement} of all the phases. 

In general, these phases can be
written as $\theta_i = \delta_i + \phi_i$,
where $\delta_i$ is a CP conserving and $\phi_i$ is a CP violating phase, 
respectively. 
Obviously, in this way it is not possible  to separate 
$\delta_i$ from $\phi_i$ because only their sum is measured.

Nevertheless, now consider the CP conjugated decay $B^- \to \pi^-\pi^-\pi^+$. 
The phase of each partial amplitude changes to $\bar{\theta}_i = \delta_i - \phi_i$.

If one then applies the fitting procedure to the DP's corresponding to both 
$B^+$ 
and $B^-$ decays, {\it one gets a direct measurement of the CP violating phase},
\begin{equation}
\phi_i = (\theta_i - \bar{\theta}_i ) /2 \hskip 0.1 cm .
\label{cpvphase}
\end{equation}

This procedure does not require  to make any assumption about FSI as other
 methods to measure CP violation demand\cite{review}. 
In B meson decays, FSI are usually assumed to be small \cite{fsi}, but this is 
not necessarily correct\cite{donoghue,gerard}. 
Here, strong phases $\delta_i$ do not 
have to be known a priori. Moreover, using this procedure one can also obtain 
a direct measurement of the strong phases
$\delta_i=({\theta}_{i} + \bar{\theta}_{i})/2$. This by-product of our method  could 
be an interesting input to other methods.

For this procedure to apply one needs at least two intermediate channels with 
different CP violating phases. Indeed, 
if all the intermediate channels have the same CP violating phase it would 
factors out; in other words, there would be no interference to peak CP 
violation.

In the decay $B^\pm \to \pi^\pm \pi^+ \pi^-$, the ${\chi_c}_0\pi^\pm$ partial 
channel 
produces the necessary interference to extract the angle $\gamma$. This
 channel  is driven by the CKM coefficients 
$V_{bc}V^*_{cd}$ and it thus has no CP violating phase. On the other side, 
the direct non-resonant contribution as well as the other resonant channels 
--- $\rho^0\pi^\pm, f_0\pi^\pm$, etc --- proceed via the CKM coefficients 
$V_{ub}V^*_{ud}$ and their amplitudes thus contain the weak phase $\gamma$. 

Unfortunately, all the partial channels but ${\chi_c}_0\pi$ contain 
also penguin diagrams which are driven by another CP violating angle, $\beta$. 
These penguin contributions are expected to be small but not 
negligible\cite{review,TP}.  

In the following, we will present our simulations of the $B^\pm$ decays.
As a first step, we will not include  penguin contributions. 
The corrections due to their inclusion 
will be studied later in this letter. They are the {\it only} source of 
theoretical uncertainties within the method presented here.

The experimental simulation consisted in the following. First, we have
generated a sample of $B^+\to \pi^+\pi^+\pi^-$ events using Monte Carlo 
technique. The dynamics was given by the function
$\cal{F}$ of Eq. (\ref{fit}), with a given set of input
parameters $a_i$ and $\theta_i$. 
Then, we fitted the generated distribution of events in the DP using the 
maximum-likelihood
fitting technique and MINUIT package\cite{MINUIT}. The fitting function
was $\cal{F}$ but now $a_i$ and $\theta_i$ are the floating 
parameters which are obtained from the fit. These two steps were then 
repeated for the CP conjugated decay $B^-\to\pi^-\pi^-\pi^+$.

In fact, we have considered a number of sets of input parameters $a_i$ 
and $\theta_i$ corresponding to various possible scenarios for  the 
unknown quantities 
involving this decay: the relative weight of each partial channel, their 
relative strong phases, and the angle $\gamma$. 

In Tables \ref{tab1} and \ref{tab2} we show 
the result of one of our simulations of the decay. 
It describes a probable scenario according to our present knowledge:

1) BR($B^+ \to {\chi_c}_0\pi^+) \sim 5 \times 10^{-5}$\cite{pap3}; 
BR(${\chi_c}_0 \to \pi^+\pi^-) \sim$ 0.8\% \cite{pdg};  
BR($B^+ \to \pi^+ \pi^+ \pi^-)_{NR}$ (non resonant) 
$\sim 10^{-5}$\cite{deshp}; 
BR($B^+\to\rho^0\pi^+$) $\sim$ $8\times 10^{-6}$\cite{kramer}; 
BR($\rho^0\to \pi^+\pi^-) \sim $100\%.  
We have assumed BR($B^+\to f_0\pi^+) \sim $BR($B^+\to \rho^0\pi^+$). 
From the square roots of the numbers above, one simply gets the 
coefficients $a_i$ of the column ``input'' in Tables \ref{tab1} and 
\ref{tab2}. 

2) The CP violating phase $\phi_2=\phi_3=\phi_4=\gamma$ for $B^+$ has been 
chosen as\cite{289} 65$^{o}$. The unknown strong FSI phases $\delta_i$ 
have been arbitrarily taken as 
$5^o, 15^o$ and $-10^o$ for NR, $f_0\pi^\pm$ and $\rho^0\pi^\pm$, 
respectively. With these numbers one 
gets the values $\theta_i$ of the column ``input'' in Tables \ref{tab1} and 
\ref{tab2}.

The NR distribution has been considered flat. We fixed the ${\chi_c}_0\pi$ 
parameters $a_1=1.0$ to have an overall normalization and $\theta_1=0$ 
to fix our phase definition.

We show in Table \ref{tab1} the result of the simulation for the $B^+$ 
decay for three different numbers 
of generated events (200, 500 and 1000). In Table \ref{tab2} 
we present the same  results for the CP conjugated decay. 

\begin{table}[tb]
\begin{center}
  \begin{tabular}{|c c|c|c|c|c|}
    \hline
decay & & input & 200 events & 500 events & 1000 events \\
    \hline
${\chi_c}_0\pi^+$ & $a_1$ & 1.0 & fixed & fixed & fixed \\

& $\theta_1$ & 0$^o$  & fixed & fixed & fixed \\
 \hline
NR & $a_2$ & 4.0 & $3.1 \pm 0.7$ & $4.1 \pm 0.6$  & $3.8 \pm 0.5$ \\
 & $\theta_2$ & $70^o$ & $(64 \pm 25)^o$ & $(71 \pm 17)^o$ &$
(66 \pm 11)^o$ \\
\hline
$f_0\pi^+$ & $a_3$ & 2.5 & $2.5 \pm 0.6$ & $2.8 \pm 0.4$ & $
2.5 \pm 0.3$ \\
& $\theta_3$ & 80$^o$  & $(99 \pm 28)^o$ & $(88 \pm 18)^o$ &$
(75 \pm 12)^o$ \\
\hline
$\rho^0\pi^+$ & $a_4$ & 3.0 & $2.1 \pm 0.6$ & $3.4 \pm 0.5$ &$
3.0 \pm 0.4$ \\
& $\theta_4$ & 55$^o$  & $(74 \pm 26)^o$ & $(39 \pm 18)^o$ &$
(50 \pm 12)^o$ \\
 \hline
  \end{tabular}
  \caption[]{Fitting results for $B^+$ Monte Carlo sample.}
  \label{tab1}
\end{center}
\end{table}

\begin{table}[tb]
\begin{center}
  \begin{tabular}{|c c|c|c|c|c|}
    \hline
decay & & input & 200 events & 500 events & 1000 events \\
    \hline
${\chi_c}_0\pi^-$ & $a_1$ & 1.0 & fixed & fixed & fixed \\

& $\bar{\theta_1}$ & 0$^o$  & fixed & fixed & fixed \\
 \hline
NR & $a_2$ & 4.0 & $4.2 \pm 1.0$ & $3.6 \pm 0.6$  & $4.2 \pm 0.4$ \\
 & $\bar\theta_2$ & $-60^o$ & $(-72 \pm 28)^o$ & $(-64 \pm 16)^o$&$
(-66 \pm 10)^o$\\
\hline
$f_0\pi^-$ & $a_3$ & 2.5 & $2.6 \pm 0.7$ & $2.4 \pm 0.6$ & $
2.4 \pm 0.4$\\
& $\bar\theta_3$ & $-50^o$  & $(-48 \pm 29)^o$ & $(-55 \pm 18)^o$
&$(-48 \pm 12)^o$\\
\hline
$\rho^0\pi^-$ & $a_4$ & 3.0 & $3.0 \pm 0.7$ & $2.8 \pm 0.7$ &$
3.3 \pm 0.5$ \\
& $\bar\theta_4$ & $-75^o$  & $(-81 \pm 27)^o$ & $(-81 \pm 16)^o$&$
(-83 \pm 11)^o$\\
 \hline
  \end{tabular}
  \caption[]{Fitting results for $B^-$ Monte Carlo sample.}
  \label{tab2}
\end{center}
\end{table}

One then uses Eq. (\ref{cpvphase}) for the three CP phase changing channels 
NR, $f_0\pi$ and $\rho\pi$. One gets
$$
\gamma=68\pm 19 \hskip 0.2 cm , \hskip 0.4 cm \gamma=68\pm 12 \hskip 0.2 cm , 
\hskip 0.4 cm \gamma=66\pm 7 \hskip 0.4 cm (NR)
$$
$$\gamma=73\pm 20 \hskip 0.2 cm , \hskip 0.4 cm \gamma=71\pm 12 \hskip 0.2 cm 
, \hskip 0.4 cm \gamma=62\pm 8 \hskip 0.4 cm (f_0\pi)
$$
\begin{equation}
\gamma=77\pm 19 \hskip 0.2 cm , \hskip 0.4 cm \gamma=60\pm 12 \hskip 0.2 
cm , \hskip 0.4 cm \gamma=66\pm 8 \hskip 0.4 cm (\rho\pi)
\label{gamma}
\end{equation}
for 200, 500 and 1000 generated events, respectively. The errors in Eq. 
(\ref{gamma}) have been obtained summing in quadrature the independent 
errors of $B^+$ and $B^-$ fits. 

These results correspond to the scenario shown in Tables I and II. 
Nevertheless, as this scenario is based in particular assumptions we 
have performed a systematic study of these results, allowing a 
{\it large variety} of other scenarios. 

First, we 
have varied the BR of the partial channels -- i.e., the square of the 
input coefficients $a_i$ -- by as much as a factor of 5 and we got acceptable 
fits with similar errors. Second, we have tried other values of the CP 
conserving phases $\delta_i$ and we got the same accuracy for any value 
of the phases, even when they were all set to zero. Third, we 
have tried many different values of $\gamma$ 
between 0 and 2$\pi$ and we have always found the {\it same accuracy} in 
the results. Finally, we have made simulations releasing the shape of 
the function $F_2$ describing the non-resonant channel 
\cite{prl-dalitz} and found no important variations in the errors of 
Eq. (\ref{gamma}). 

We are then confident that in any acceptable scenario for this decay, 
the error to extract the angle $\gamma$ would be similar.
This procedure thus brings a simple way of predicting {\it the error} in the 
measurement of $\gamma$; it would only depend on the number of reconstructed 
events --- this is not the case for many other methods\cite{BaBar}.

It is worth mentioning another important point of our simulations. The method 
has no discrete ambiguities; accordingly, we always get only one value of 
$\gamma$ from the fit.

Let us now discuss the real scenario including penguins, studying by how much 
their inclusion modifies our previous results. For example, in the $f_0\pi^+$ 
channel the measured quantity $a_{3} e^{i\theta_3}$ is in fact
\begin{equation}
a_{3} e^{i\theta_3} = T e^{i(\delta_T+\gamma)} + P e^{i(\delta_P-\beta)} 
\hskip 0.1 cm ,
\label{penguin}
\end{equation}
where $T e^{i(\delta_T+\gamma)}$ is the tree contribution and 
$P e^{i(\delta_P-\beta)}$ is the penguin one\cite{review}; 
$\delta_T$ and $\delta_P$ are the strong phases.

A pictorial representation of Eq. (\ref{penguin}) is shown in Figure 1. 
It shows that when measuring the angle $\theta_3$ we are missing the 
actual tree phase by an angle $\epsilon_+$. The same argument holds 
for $B^-$, leading to an angle $\epsilon_-$. As a result, Eq. (\ref{cpvphase}) 
becomes
\begin{equation}
\phi_{3} = (\theta_{3} - \bar{\theta}_{3} ) /2 = 
\gamma +(\epsilon_++\epsilon_-)/2 \hskip 0.1 cm .
\label{fin}
\end{equation}
Thus, $\epsilon\equiv (\epsilon_++\epsilon_-)/2$ is the theoretical error 
of our method. Figure 1 shows that the worst case corresponds to the 
configuration when the tree and the penguin contributions
are orthogonal in the complex plane. As a consequence, we have
\begin{equation}
|\epsilon_{\pm}| \leq \arctan(P/T) \hskip 0.1 cm .
\label{epsilon}
\end{equation}

\begin{figure}
\begin{center}
\mbox{\epsfig{figure=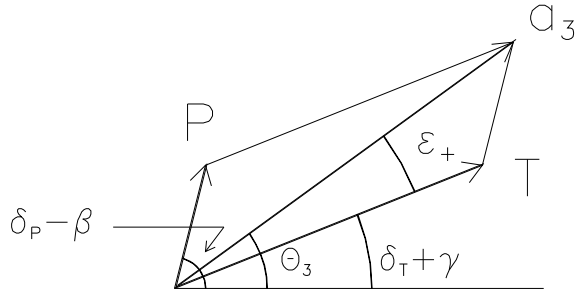,width=8cm}}
\caption{{\it Tree, penguin, and the measured contributions.}}
\end{center}
\protect\label{Fig1}
\end{figure}

The actual value of the ratio $P/T$ is not known at present. An estimate was 
obtained for the decay $B\to \pi\pi$; $P/T \sim 0.2$ \cite{TP}. In our case, 
one expects the 
ratio $P/T$ to be of the same order. 
Assuming $P/T = 0.2$, the uncertainty on $\gamma$ 
extraction due to penguin contribution would be at most $\sim 11^o$. 
Anyway, Eq. (\ref{epsilon}) shows that as long as $P/T$ remains not very large, 
the penguin pollution does not invalidate the method; for example even for the 
improbable value $P/T \sim 0.5$, $\epsilon \leq 26^o$. 
 
The inclusion of final state rescattering does not spoil our analysis of 
the error. The  $B^+ \to f_0\pi^+$ decay proceeds through a unique 
isospin amplitude and thus\cite{gerard} Eq. (\ref{penguin}) remains 
unchanged. For partial decays with more than one isospin amplitude --- 
as $B^+\to \rho^0\pi^+$ for example ---  
 the form of Eq. (\ref{penguin}) does not change either, but the {\it
interpretation} of the two terms in this equation does. Indeed, even
with rescattering one still has two types of diagrams: the first type
having weak phase $\gamma$, the second having weak phase
$-\beta$. But now, for example, the term we call $T$ in 
Eq. (\ref{penguin}) would be a more subtle combination of tree, 
color-suppressed and annihilation quark diagrams\cite{gerard}. Thus,
the method also applies to intermediate channels with more than one isospin 
amplitude; nevertheless, a complete isospin analysis 
would be required to stablish the theoretical uncertainty  --- note 
however that this is only necessary if rescattering effects are 
found to be large.

Let us now study the experimental feasibility of this method. Using Eqs. 
(\ref{gamma}) and (\ref{fin}-\ref{epsilon}) one can immediately get the error 
in the extraction of $\gamma$ according to the number of reconstructed events. 
For example, with 1000 reconstructed events and $P/T = 0.2$, this method 
would give $\gamma$ with a statistical plus theoretical error of 
22$^o$ with 90\% of CL. Moreover, 
combining the measurement of $\gamma$ using all the intermediate channels 
NR, $\rho\pi$ and $f_0\pi$ would certainly decrease the error of this 
procedure.

Much less statistics is needed to simply {\it detect} CP violation in this 
decay. One observes CP violation 
when, e.g., $\bar{\theta}_2 - \theta_2$, is different from zero. Assuming  $\gamma \sim 65^o$ then from Eq. (\ref{gamma}) one concludes 
that only 200 events are needed in order to detect a 3$\sigma$ CP violation 
effect.

Three main features allows us to be optimistic about the possibility of 
doing this analysis in a short period of time, either in BaBar, KEK or 
CLEOIII. First, tagging is not required because one needs only charged 
B's. Second, the three detected particles are charged, thus we expect 
the efficiency to be high and the background to be not very large. 
Third,  the method itself demands small statistics. For the 
$B^+\to \pi^+\pi^+\pi^-$ decay, assuming for example a total BR 
of $3\times 10^{-5}$ and a reconstruction efficiency of 60\% \cite{BaBar} 
one would expect to need about 2 years of running of BaBar to reconstruct 1000 
events. Of course, a full experimental simulation of this decay is 
required to have definite conclusions.

In the end, let us summarize the main points of the procedure to extract 
$\gamma$ presented in this letter. This method brings a direct measurement 
of the CP violating angles. No previous knowledge of BR's or FSI phases 
is required. No necessity of making complicated triangle constructions 
is needed. 
The Dalitz analysis deals directly with amplitudes; thus, it is 
{\it linearly} sensitive to suppressed decays, as $\chi_c\pi$. Because of 
all of this, it is natural that this method does provide a measurement 
which demands less statistics than other methods. Finally, as one directly 
gets the angle itself --- instead of, e.g. twice its sine or its cosine as 
in other methods --- there are no discrete ambiguities.

The limitations of this method are mainly due to the fact that in order to 
have a complete knowledge of the errors in the extraction of $\gamma$ one 
needs the ratio $P/T$. One hopes that in the near future this ratio will 
be better known both from the theoretical and the experimental sides.
As a result, although this method will probably not yield a very accurate 
measurement of $\gamma$, it will certainly allow a simple and effective 
step to have a quick constraint in the value of this angle.

As a by-product of this method to extract $\gamma$, we have presented in 
this letter a general procedure to measure CP violating angles in charged 
three body decays. It is a natural and clear method. The procedure is 
quite general as it applies to any three body decay. Moreover, the whole 
procedure applies for {\it any CP violating phase}. For example, using 
this methods with existing data from D meson decays one could obtain 
upper limits of less than 1$^o$ for many CP violating angles. This could 
be used to constrain beyond the Standard Model physics.

\end{document}